\documentclass[11pt]{article}
\usepackage{graphicx}

\newcommand{\BaBarYear}       {00}
\newcommand{\BaBarNumber}     {11}
\newcommand{\SLACPubNumber} {8622}
\newcommand{\BaBarType}     {PROC}  

\usepackage{relsize}
\def\babar{\mbox{\slshape B\kern-0.1em{\smaller A}\kern-0.1em
    B\kern-0.1em{\smaller A\kern-0.2em R}}}

\def\en         {\ensuremath{e^-}}      
\def\ep         {\ensuremath{e^+}}
  
\def\epem       {\ensuremath{e^+e^-}}

\def\Kbar  {\kern 0.2em\overline{\kern -0.2em K}{}}

\def\KS    {\ensuremath{K^0_{\scriptscriptstyle S}}}

\def\Kzb   {\ensuremath{\Kbar^0}}
\def\KzKzb {\ensuremath{K^0 \kern -0.16em \Kzb}}
\def\Dz    {\ensuremath{D^0}}
\def\Dbar  {\kern 0.2em\overline{\kern -0.2em D}{}}

\def\Dzb   {\ensuremath{\Dbar^0}}
\def\DzDzb {\ensuremath{D^0 {\kern -0.16em \Dzb}}}
\def\Dstar   {\ensuremath{D^*}}

\def\Bz    {\ensuremath{B^0}}
\def\B     {\ensuremath{B}}
\def\Bbar  {\kern 0.18em\overline{\kern -0.18em B}{}}

\def\Bzb   {\ensuremath{\Bbar^0}}

\def\BzBzb {\ensuremath{B^0 {\kern -0.16em \Bzb}}}

\def\jpsi  {\ensuremath{{J\mskip -3mu/\mskip -2mu\psi\mskip 2mu}}} 
\def\psitwos {\ensuremath{\psi{(2S)}}}
\mathchardef\Upsilon="7107
\def\Y#1S{\ensuremath{\Upsilon{(#1S)}}}

\def\FourS {\Y4S}

\mathchardef\Deltares="7101
\mathchardef\Xi="7104
\mathchardef\Lambda="7103
\mathchardef\Sigma="7106
\mathchardef\Omega="710A
\def\Deltabar   {\kern 0.25em\overline{\kern -0.25em \Deltares}{}}
\def\Lbar {\kern 0.2em\overline{\kern -0.2em\Lambda\kern 0.05em}\kern-0.05em{}}
\def\Sigbar{\kern 0.2em\overline{\kern -0.2em \Sigma}{}}
\def\Xibar{\kern 0.2em\overline{\kern -0.2em \Xi}{}}
\def\Obar{\kern 0.2em\overline{\kern -0.2em \Omega}{}}
\def\Nbar{\kern 0.2em\overline{\kern -0.2em N}{}}
\def\Xbar{\kern 0.2em\overline{\kern -0.2em X}{}}

\def\bpsiks{\ensuremath{B^0 \to \jpsi \KS}}

\def\ev   {\ensuremath{\rm \,e\kern -0.08em V}}
\def\kev  {\ensuremath{\rm \,ke\kern -0.08em V}} 
\def\mev  {\ensuremath{\rm \,Me\kern -0.08em V}} 
\def\gev  {\ensuremath{\rm \,Ge\kern -0.08em V}} 
\def\gevc {\ensuremath{{\rm \,Ge\kern -0.08em V\!/}c}} 
\def\tev  {\ensuremath{\rm \,Te\kern -0.08em V}}
\def\mevc {\ensuremath{{\rm \,Me\kern -0.08em V\!/}c}} 
\def\gevcc{\ensuremath{{\rm \,Ge\kern -0.08em V\!/}c^2}} 
\def\mevcc{\ensuremath{{\rm \,Me\kern -0.08em V\!/}c^2}}

\def\cm   {\ensuremath{\rm \,cm}}

\def\mum  {\ensuremath{\,\mu\rm m}} 

\def\invpb {\ensuremath{\mbox{\,pb}^{-1}}}

\def\invfb   {\ensuremath{\mbox{\,fb}^{-1}}}
\def\mus  {\ensuremath{\rm \,\mus}}

\def\cms         {\ensuremath{{\rm \,cm}^{-2} {\rm s}^{-1}}}  
\def\mus        {\ensuremath{\,\mu{\rm s}}}    
\def\gsim{{~\raise.15em\hbox{$>$}\kern-.85em
          \lower.35em\hbox{$\sim$}~}}
\def\lsim{{~\raise.15em\hbox{$<$}\kern-.85em
          \lower.35em\hbox{$\sim$}~}}

\def\CP                 {\ensuremath{C\!P}}

\def\to                 {\ensuremath{\rightarrow}}

\def\pep2{PEP-II}
\def\BF{$B$ Factory}
\def\abf {asymmetric \BF}

\newcommand{\dedx}{\ensuremath{\mathrm{d}\hspace{-0.1em}E/\mathrm{d}x}}

\def\sb{${\sin\! 2 \beta   }$}

\providecommand{\eqref}[1]{Eq.~(\ref{eq:#1})}

\def\jetset74   {\mbox{\tt Jetset \hspace{-0.5em}7.\hspace{-0.2em}4}}

\def\bpsik{\ensuremath{B^+ \to \jpsi K^+}}
\def\bpsitwok{\ensuremath{B^+ \to \psitwos K^+}}
\def\bpsitwoks{\ensuremath{\Bz \to \psitwos \KS}}
\def\bdstar{\ensuremath{\Bz_d \to D^{*-}l^+\nu}}

\long\def\inst#1{\par\nobreak\kern 4pt\nobreak
    {\it #1}\par\vskip 10pt plus 3pt minus 3pt}

\begin{document}
{\pagestyle{empty}

\begin{flushright}
\begin{flushright}
\babar-\BaBarType-\BaBarYear/\BaBarNumber \\
SLAC-PUB-\SLACPubNumber
\end{flushright}
\end{flushright}

\par\vskip 4cm

\begin{center}
First Year of the \babar\ Experiment Data Taking
\end{center}
\bigskip

\begin{center}
\large 
Marc Verderi

{\it Laboratoire de Physique Nucl\'eaire des Hautes Energies

Ecole Polytechnique

91128 Palaiseau, FRANCE}

(for the \babar\ Collaboration.)

\mbox{ }\\
\end{center}
\bigskip \bigskip

\begin{center}
\large \bf Abstract
\end{center}
The \babar\ experiment at the \pep2\ \abf\ observed its first hadronic events on the $26^{th}$ of
May 1999. We present the progress made so far on the data taking, the \babar\ detector
performances and the physics analyses.\\

\vfill
\begin{center}
Contribued to the Proceedings of the XX$^{th}$ Physics in Collision, \\
6/29/2000---7/1/2000, Lisbon, Portugal
\end{center}

\vspace{1.0cm}
\begin{center}
{\em Stanford Linear Accelerator Center, Stanford University, 
Stanford, CA 94309} \\ \vspace{0.1cm}\hrule\vspace{0.1cm}
Work supported in part by Department of Energy contract DE-AC03-76SF00515.
\end{center}
}

\setcounter{footnote}{0}

\section{Introduction}
\label{sec:Introduction}
 
The primary goal of the \babar\ experiment at \pep2\ is to over-constraint the unitarity
triangle. The sides of this triangle can be measured through non-\CP\  violating physics, while its
angles are accessible through \CP-violating processes \cite{ref:babarphys}. The first measurable angle will
be \sb\ with the so called ``golden plated'' channel \bpsiks\ . Because of the variety of processes
to be studied, \babar\ has been designed as a general HEP detector \cite{ref:babartech}  and can also be useful for non-CKM
matrix physics.\\

The \CP\ violation is studied through time-dependant \CP\ asymmetries:

\begin{equation}
A_{CP}(t) = \frac
	{\Gamma(\Bz \rightarrow f_{CP})-\Gamma(\Bzb \rightarrow f_{CP})}
	{\Gamma(\Bz \rightarrow f_{CP})+\Gamma(\Bzb \rightarrow f_{CP})}
	 =  A_{CP}.\sin(\Delta m_B\cdot t) \label{eq:asymcp}
\end{equation}\\[-0.60cm]

where $f_{CP}$ denotes a particular \CP\ eigenstate final state and $\Delta m_B$ the mass difference between the two
neutral $B$ mass eigenstates. In the Standard Model, this $A_{CP}$ term for the \bpsiks\ channel is directly \sb .


The measurement of $A_{CP}(t)$ relies on the determination of two ingredients: the \B\ meson decay time $t$
and the flavor \Bz\ or \Bzb\ of this \B\ meson at $t=0$.\\

The principle adopted by asymmetric B Factories is to produce \BzBzb\ pairs through the asymmetric
\epem collisions, $\epem \rightarrow \FourS \rightarrow \BzBzb$,
so that the \BzBzb\ system formed is boosted.

Because the \FourS\ is $J^{PC} = 1^{--}$, the \BzBzb\ system is in an antisymmetric state.
That garanties that at all times, the two \B s will have opposite flavors. Once one of the \B\ decays
it is possible to infer its flavor by measuring its decay products. The sign of a lepton present in
the decay products, taking
care if this lepton is direct or indirect, can be used, as well as kaon sign, for example.
This flavor determination is called the ``tagging''. Thanks to the antisymmetry of the \BzBzb\ system,
we know that the other \B\ meson, has the opposite flavor at that time, which define $t = 0$ in ~\ref{eq:asymcp}.

The measurement of the distance $\Delta z$ between the two decay vertices,
which is about 260 \mum\ on average at \pep2,
provides the measurement of the $\B\rightarrow f_{CP}$ decay time $t$, the second ingredient of equation ~\ref{eq:asymcp}, through
$t = \Delta z/\beta\gamma c$.\\

Beyond the need of an asymmetric machine, other requirements have to be met: the \B\ vertices
reconstruction implies an excellent tracking and vertexing capability. Because $\B \rightarrow f_{CP}$ modes
have low branching ratios, typically at the $5\times 10^{-5}$ level, a high luminosity is needed. The
tagging requires good lepton and kaon identification.\\

Those requirements have driven the \pep2\ and \babar\ designs.

\section{The \pep2\ machine}

\begin{figure}[!htb]
\begin{center}
\includegraphics[height=7cm]{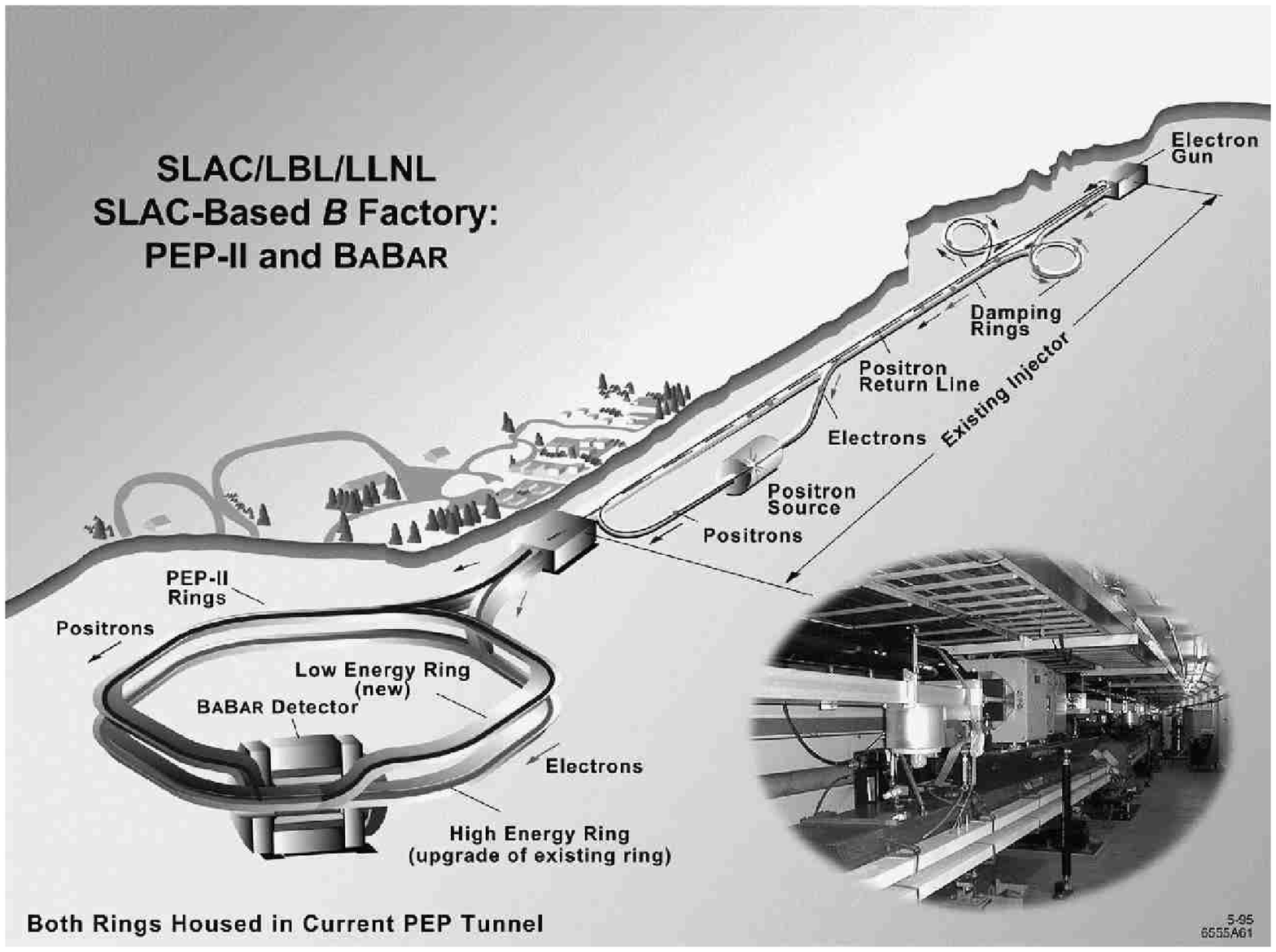}
\caption{The \pep2\ storage rings.}
\label{fig:pep2}
\end{center}
\end{figure}

The \pep2\ machine reuses the existing Linac injector part (see figure \ref{fig:pep2}). The
positrons circulate in the Low Energy Ring (LER) which was also an existing part.
The High Energy Ring (HER) has been specially
built.

The beam energies and center of mass boost are:
\[\en (9 \gev) \otimes \ep (3.1 \gev) \Rightarrow \beta\gamma \sim 0.56\]

\begin{table}[!htb]
\caption{\pep2\ performances}
\begin{center}
\begin{tabular}{||c|c|c|c||}  \hline
			&	design			&	achieved		&	typical			\\ \hline
Luminosity (\cms)	&	$3\times 10^{33}$	&	$2.2\times 10^{33}$	&	$1.9-2.1\times 10^{33}$	\\ \hline
LER current ($A$)	&	2.14			&	1.7			&	1.1			\\
LER lifetime		&	$4h@2A$			&	$3.5h@1A$		&	$3h@1A$			\\ \hline
HER current ($A$)	&	0.75			&	0.92			&	0.65			\\
HER lifetime		&	$4h@1A$			&	$11h@0.9A$		&	$9h@0.65A$		\\ \hline
\#bunches(HER/LER)	&	1658			&	1658			&	829			\\ \hline
\end{tabular}
\end{center}
\label{tab:peperf}

\end{table}
Table ~\ref{tab:peperf} shows the designed, achieved and typical values of the \pep2\ parameters. The designed values have all been
or are close to be achieved, and are also close to the typical ones.\\

\begin{figure}[!htb]
\begin{center}
\includegraphics[height=6cm]{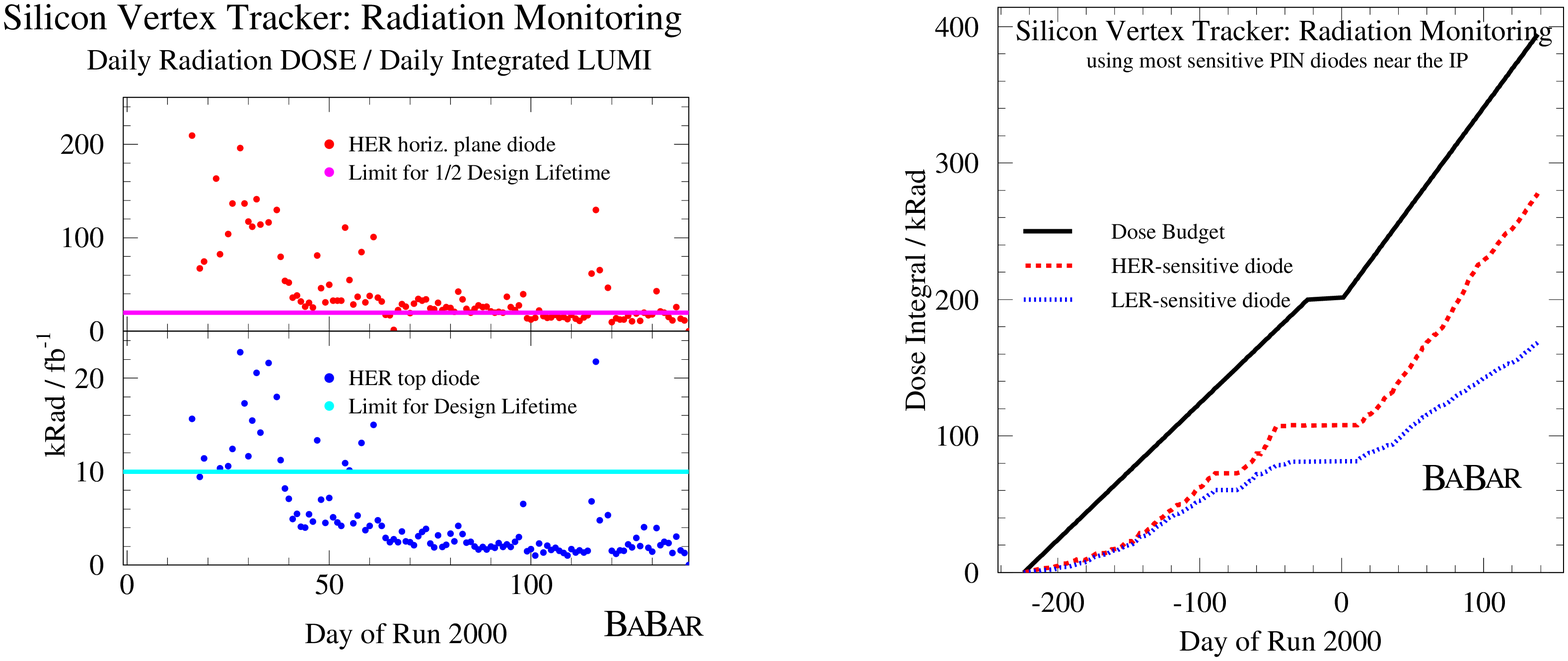}\\[-0.6cm]
\caption{Silicon Vertex Tracker daily delivered dose (left) and dose budget (right).}
\label{fig:doses}
\end{center}
\end{figure}

With such high currents, one may worry about the level of radiation delivered to the \babar\ subsystems.
Figure \ref{fig:doses}(left) shows the daily radiation dose delivered by the HER at two positions in the Silicon Vertex Tracker,
in the horizontal (top) and vertical (bottom) planes over the past year. Limits for half and full design lifetimes
are shown on this figure.
Those dose rates can be turned into dose budgets which can be allowed. These dose budgets are drawn on
figure \ref{fig:doses}(right), showing that
they are not overspent.\\[-0.5cm]

The \babar/\pep2\ ``Factory Mode'' can be illustrated with a few numbers: the daily recorded luminosity is at present in the 
120 to 140 \invpb\ range, \pep2\ delivers long duration fills, the record being 22 hours, and the \babar\ data taking
efficiency is typically $95$ to $97 \%$.\\

The accumulated data sample at the time of this conference is about $11\invfb$.

\section{\babar\ detector highlights}

\begin{figure}[!htb]
\begin{center}
\includegraphics[height=6cm]{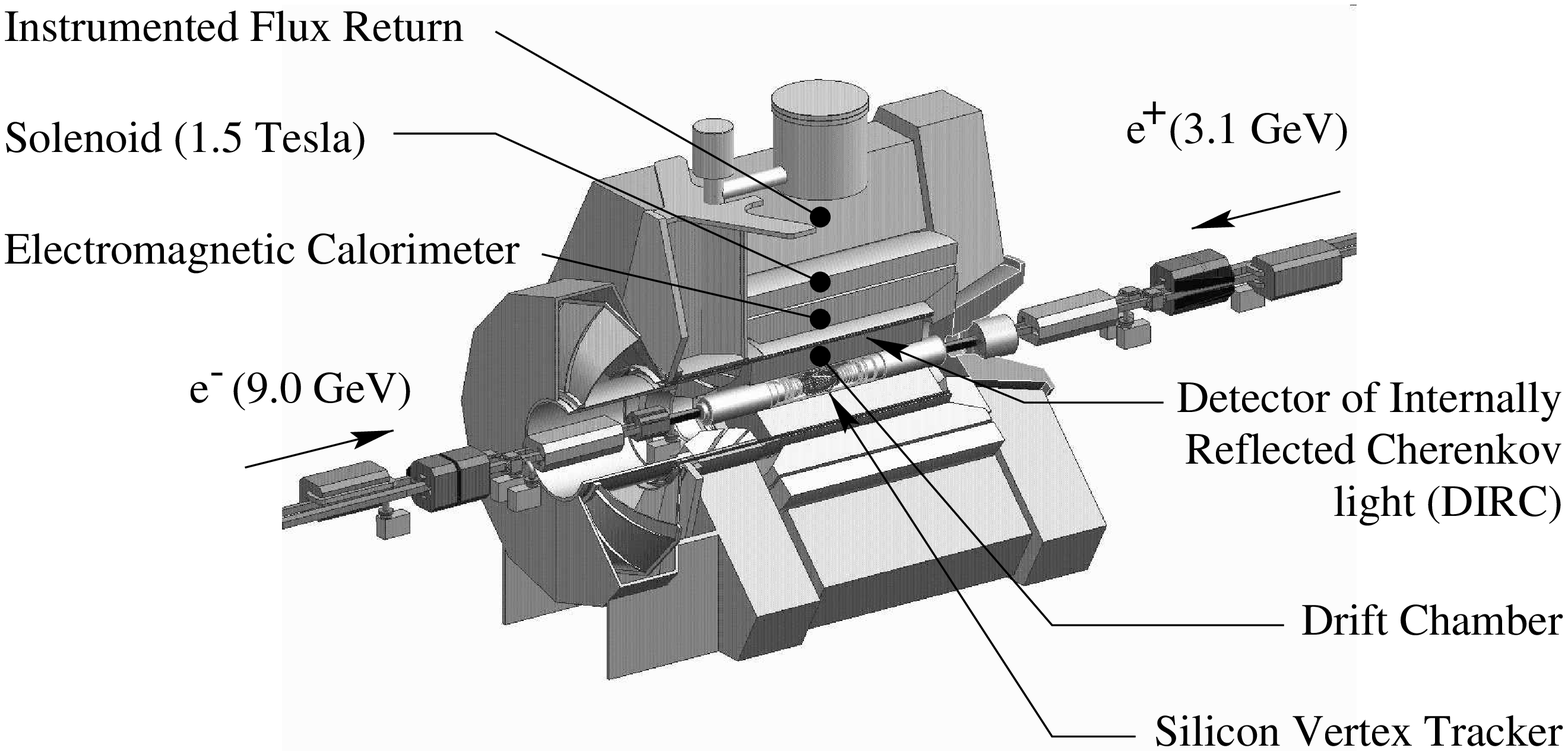}
\caption{\babar\ detector layout.}
\label{fig:babar}
\end{center}
\end{figure}

Figure \ref{fig:babar} shows the \babar\ layout. Electrons enter from the front \babar\ side.
The center of mass boost induced the asymmetric design with an interaction point displaced from the geometrical center
in the backward direction.

\subsection{Silicon Vertex Tracker (SVT)}

The SVT is made of 5 layers, double side silicon. It is built in radiation-hard technologies, allowing up to 2 Mrad dose.
The SVT extends from $3.3\ \cm$ to $14.6\ \cm$ in radius. The three inner layers are dedicated to measurement of the
impact parameter $d_0$ and the angle of the tracks.
The two outer layers allow track pattern recognition of slow particles,
which makes the Silicon Vertex ``Tracker'' capable of {\em standalone tracking}. The $z$ and $d_0$ resolutions
have been measured using muons of transverse momentum above 2 \gev to be better than $40\ \mum$.

The alignment of the SVT is an important issue. The relative SVT/Drift Chamber diurnal motions are measured to be
as large as $70\ \mum$. The alignment corrections are computed on a run by run basis as part
of the reconstruction, in a procedure called the {\em ``rolling calibration''}.

\subsection{Drift Chamber (DCH)}

The Drift Chamber is made of $\sim$ 7000 drift cells with hexagonal field wire pattern. It is filled with a 80/20 \% 
He/C$_4$H$_{10}$ gas mixture in order to minimize the multiple scattering.

The single hit resolution measured with tracks above 1 \gev\ is approximately 125 \mum, better
than the design value ($\sim 140\ \mum$). The \dedx\ resolution is about 7.5 \%
for Bhabha events (the design value being 7 \%).
This \dedx\ resolution allows a $2\sigma \ K/\pi$ separation up to 700 \mevc.

\subsection{Detector of Internally Reflected Cherenkov light (DIRC)}

The DIRC is the main particle identification device of \babar . It provides identification of particles above 500\ \mevc .
The radiator part is made of quartz bars arranged in 12 sectors. The
Cherenkov light produced by charged particles in the quartz is transported by internal reflections down to the Stand
Of Box, a water filled expansion, visible on figure ~\ref{fig:babar} at the rear of \babar .
An array of 10\,752 photo-multipliers reads the image formed by the Cherenkov light.

The Cherenkov angle resolution per track is $\sigma(\theta_C)$ = 3.0 mrad,
for a targeted resolution of 2.0 mrad. This resolution allows presently
a $2.1\sigma \ K/\pi$ separation at 4 \gev .

\begin{figure}[!htb]
\begin{center}
\includegraphics[height=6cm]{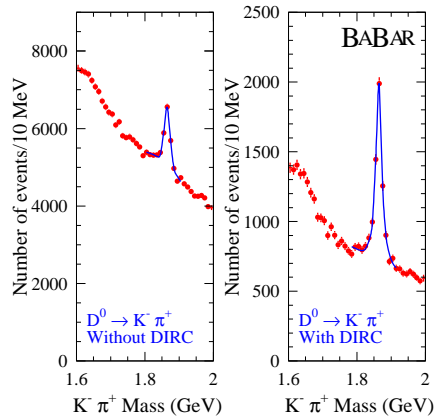}\\[-0.5cm]
\caption{$D^0\rightarrow K^\pm\pi^\mp$ signal peaks where the $K^\pm$ is inside the DIRC acceptance, without (left)
and with (right) kaon indentification ($\theta_C(K^\pm)$ required to be within $2\sigma$ of the kaon hypothesis). This allows a
background rejection of a factor of about 5. (Note that the remaining background contains true kaons.)}
\label{fig:dirc}
\end{center}
\end{figure}

Figure ~\ref{fig:dirc} shows a $D^0\rightarrow K^\pm\pi^\mp$ signal peaks, without (left) and with (right) kaon
identification from the DIRC.

\subsection{Electromagnetic Calorimeter (EMC)}

The Electromagnetic Calorimeter consists of 6580 CsI(Tl) crystals. It is made of a barrel and a forward end-cap. The resolution
achieved on $\pi^0\rightarrow\gamma\gamma$ signal mass peak, requiring $E_\gamma >$ 30 \mev, is 6.9 \mev, in good
agreement with Monte Carlo expectation.

Electron identification is performed in the EMC. The electron detection efficiency $\epsilon(e^\pm)$ is typically 92 \%
above 1 \gevc . The pion misidentification probability $\epsilon(\pi^\pm)$ is measured below 2 \gevc\ on a sample of pions
from $K^0_s\rightarrow\pi^+\pi^-$ decays and is found to be about 0.3 \%.



\subsection{Instrumented Flux Return (IFR)}

The Instrumented Flux Return is made of bakelite-based resistive plate chambers sandwiched between iron plates. It provides
muon identification above 500 \mevc\ and $K^0_L$ detection.

Beetwen 1.5 \gevc\ and 3 \gevc\ the muon identification efficiency is about 75 \%
and the pion misidentification probability 2.4 \%.

\section{Physics status in \babar }

Lot of activity is going on within \babar\ both on \CP\ and non-\CP\ physics. The analyses are in ``validation'' phase, which
means that the current concerns are the understanding and the control of the mass spectra resolutions, yields, and also the control
of more complex algorithm like the tagging one.\\

The idea here is not to make an exhaustive status of the analysis but rather try to give a flavor of the current effort
and point out aspects relevant for \CP\ measurement.

\subsection{$\B\rightarrow\jpsi K$ and $\B\rightarrow\psitwos K$}

\begin{figure}[!htb]
\begin{center}
\includegraphics[height=5cm]{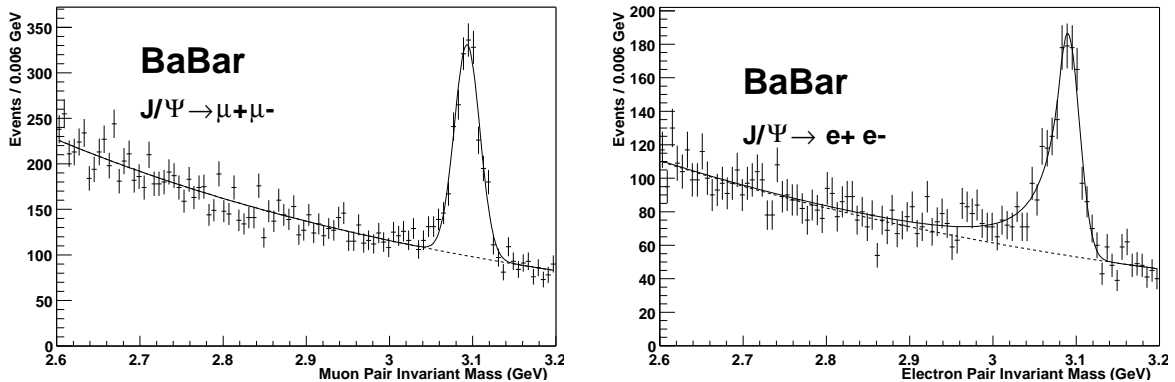}
\caption{\jpsi\ candidates in the di-muon (left) and di-electron (right) channels.}
\label{fig:psi}
\end{center}
\end{figure}

Figure ~\ref{fig:psi} shows the di-muon(left) and di-electron(right) invariant mass spectra.
The radiative tail due to bremsstrahlung is clearly visible on the di-electron mass peak.

Those spectra are obtained from a $\sim$ 2 \invfb data sample. The numbers of \jpsi\ candidates in
both peaks are compatible with expectations. The mass resolution of the $\jpsi\rightarrow\mu^+\mu^-$ signal peak
is $\sim$ 15 \mevcc . Thanks to the ``rolling calibration'' procedure mentionned above, this resolution is improving now
and is close to 11 \mevcc .\\

\begin{figure}[!htb]
\begin{center}
\includegraphics[height=10cm]{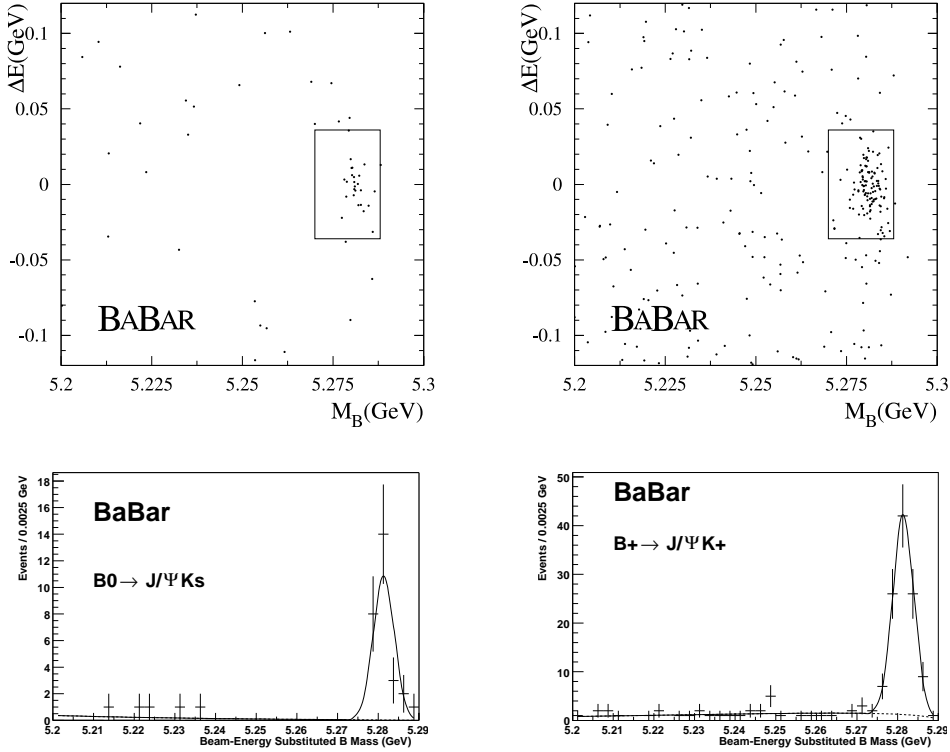}
\caption{\bpsiks\ (left) and \bpsik\ (right) candidates from a $\sim$ 2 \invfb data sample. Upper plots show the candidates
in the $\Delta E$ versus the ``beam constrained mass'' $M_B$ defined in the text. The squares on the top plots represent the
``signal region'' and correspond to roughly $\pm$ 3 $\sigma$ on $\Delta E$ and $M_B$.
Bottom plots are projections of $M_B$ within a $\pm 3 \sigma$ band in $\Delta E$.}
\label{fig:bks}
\end{center}
\end{figure}

Figure \ref{fig:bks} shows the \bpsiks\ (left) and \bpsik\ (right) candidates from a $\sim$ 2 \invfb data sample. The former
are used in the \CP\ violation measurement.
The upper plots show the signals in the ``$\Delta E$'' versus the ``beam constrained mass'' $M_B$ plan defined as follows:
\begin{eqnarray}
	M_B      & = & \sqrt{E^{*\ \ \ \ \, 2}_{Beam} - p^{*\ 2}_B} \\
	\Delta E & = & E^*_B - E^*_{Beam}
\end{eqnarray}

where all quantities are expressed in the \FourS\ center of mass, $E^*_{Beam}$ being the beam energy, $p^*_B$ and $E^*_B$ being
the measured \B\ candidate momentum and energy respectively.\\

The number of events observed in the signal regions (see figure ~\ref{fig:bks}) are 28 and 109 in
the \bpsiks\ and \bpsik\ channels respectively and are compatible with expectations.
The $M_B$ projections (see figure ~\ref{fig:bks}) show that both signals are very clean.

\bpsik\ events are also used to check on the data the reconstruction of $\Delta z$ which enters the \CP\ 
asymmetry (equation ~\ref{eq:asymcp}) through $ t = \Delta z/\beta\gamma c $. The resolution is measured
to be $\sigma(\Delta z) \sim 100\mum$ (with a 20 \%
additional component of $\sim 320 \mum$ of $\sigma$) in good agreement with Monte Carlo expectation.\\

\begin{figure}[!htb]
\begin{center}
\includegraphics[height=9cm]{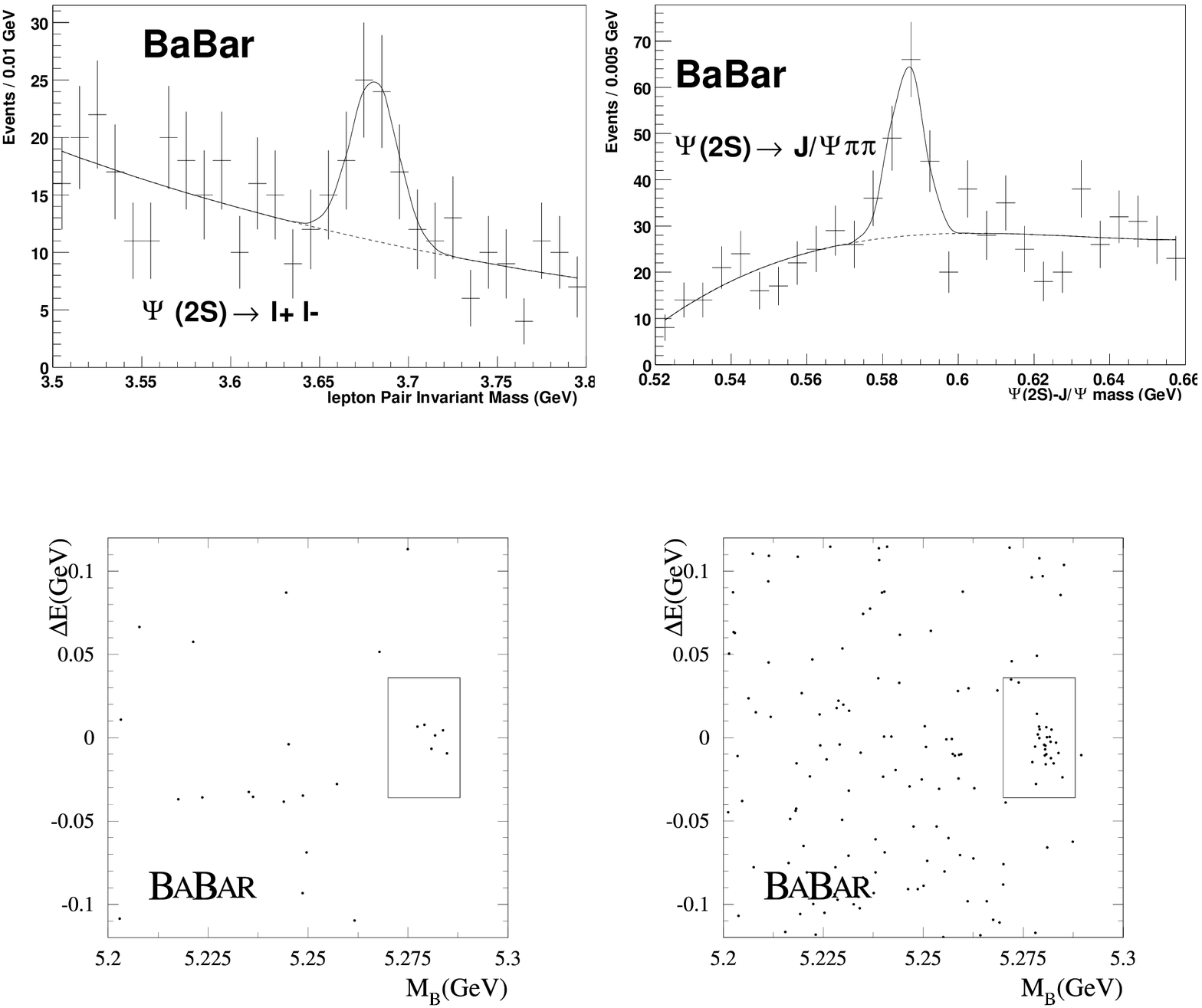}
\caption{Top: \psitwos\to $l^+l^-$ (left) and \psitwos\to\jpsi$\pi^+\pi^-$ (right) candidates. The latter are 
displayed making the difference of masses between the \psitwos\ and
\jpsi\ candidates.
Bottom: \bpsitwoks\ candidates (left), \bpsitwok\ candidates (right) formed using above \psitwos .}
\label{fig:b2ks}
\end{center}
\end{figure}

Other channels interesting for the \CP\ violation measurement are investigated.
Figure ~\ref{fig:b2ks}(bottom) shows for example the $\B\rightarrow\psitwos K$ candidates. The related \psitwos\ charmonia are
reconstructed in the \psitwos\to $l^+l^-$ and \psitwos\to\jpsi$\pi^+\pi^-$ channels ( figure ~\ref{fig:b2ks}(up)). Six events are
observed in the signal box of the \CP\ channel \bpsitwoks\  with a 0.5 estimated background and a 19 events yield is found
in the \bpsitwok\ channel.

\subsection{\B\ to open charm and \CP\ engineering}

\B\ decays to charm provide larges samples of clean reconstructed \B s in semileptonic and hadronic modes.
Beyond the intrisic physics interest, \B\ to open charm allow powerful \CP\ engineering studies.
We expose here the important issue of the control of the tagging algorithm and present the strategy
adopted by \babar\ to perform this control.\\

The tagging algorithm is characterized by two quantities: the tagging efficiency $\epsilon_{tag}$ which is the fraction
of \B s for which the tagging algorithm gives an answer, and the mis-tag fraction $w$ which is the wrong \Bz /\Bzb\ assignment
probability.

The mis-tag fraction is a key parameter because it induces two effects:
\begin{itemize}
\item It increases the uncertainty $\sigma$ on \sb :
	\[  \sigma \rightarrow \frac{\sigma}{\sqrt{\epsilon_{tag}(1-2w)^2}} =
	 \frac{\sigma}{\sqrt{\epsilon_{tag}^{effective}}}  \]
(Note that $\epsilon_{tag}^{effective}$ is typically of the order of 20 to 30 \%.)
\item But moreover it induces a {\em bias} on the asymmetry:
	\[ A_{CP} \propto \frac{A_{CP}^{observed}}{(1-2w)} \]
\end{itemize}


Since the tagging algorithm exploits mainly all the detector subsystems for leptons and kaons identification,
many possible sources of systematics errors are faced at once.

The strategy adopted by \babar\ is to perform a global control on the data themselves.

Samples of neutral \B\ mesons with known \Bz\ or \Bzb\ flavor are reconstructed. The tagging algorithm is applied on the
recoil tracks, as in a \CP\ analysis. The sample where both \B\ mesons are found to have the same flavor is retained.
Two contributions form this sample: one from \Bz -\Bzb\ mixing events and the other one from non-mixing
events, but with a wrong \Bz/\Bzb\ assignment from the tagging algorithm.
Each contribution has a well known time distribution, controled by the $\Delta m_B$ parameter. 

A fit on the time distribution on this ``same flavor events sample''
allows to separate those two contributions and thus to extract the $w$ parameter.
Moreover, by extracting $w$ on a given time window, using the knowledge on $\Delta m_B$, the $w$ value obtained can be re-used
outside this time window to measure back $\Delta m_B$ as a cross-check.\\

This study can be done using various \B\ to open charm channels. We illustrate here the statistical power of this method
with the \bdstar\ channel.\\

The \bdstar\ events are reconstructed within three channels according to the \Dz\ decay from the $\Dstar \to \Dz \pi_{soft}$ 
decay:\\[-1.2cm]

\begin{eqnarray}
\Dz &\rightarrow& K\,\pi \nonumber \\[-0.1cm]
\Dz &\rightarrow& K\,\rho(\pi\pi^0) \nonumber \\[-0.1cm]
\Dz &\rightarrow& K\,3\pi \nonumber
\end{eqnarray}

The selection is based on a missing mass to be compatible with a neutrino mass, which translates into the constraint
$ -1 < \cos(\theta_{B\,-\,D^*l}) < +1 $
and the difference of mass $\Delta m$ between the \Dstar\ and the \Dz\ candidates:
$\Delta m = m(K\pi[\pi][\pi]\pi_{soft})\,-\,m(K\pi[\pi][\pi])$
to be compatible with the nominal difference of masses.\\

\begin{figure}[!htb]
\begin{center}
\includegraphics[height=6.0cm]{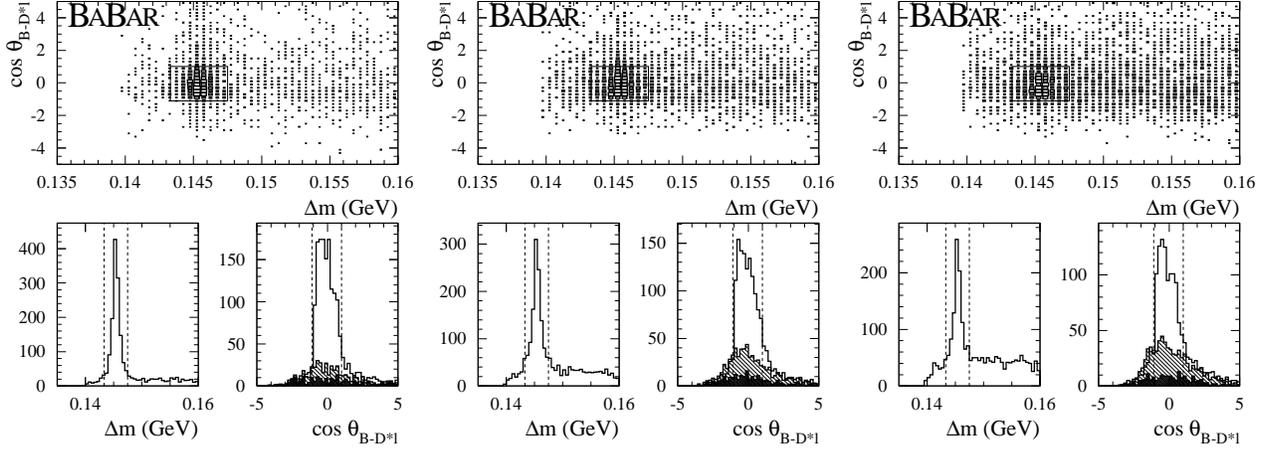}
\caption{\bdstar\ candidates. The left, middle and right plots are the $\Dz \rightarrow K\pi$, $\Dz \rightarrow K\pi\pi^0$ and
$\Dz \rightarrow K3\pi$ samples respectively. The samples are shown in the $\cos(\theta_{B\,-\,D^*l})$ versus $\Delta m$ plans
defined in the text. The 1D projections are also shown.}
\label{fig:dstlnu}
\end{center}
\end{figure}

The three \B\ samples from $\Dz \rightarrow K\pi$, $\Dz \rightarrow K\pi\pi^0$ and $\Dz \rightarrow K3\pi$ channels are
shown on figure ~\ref{fig:dstlnu}.
The luminosity used is $\sim$ 3.3 \invfb . The total number of reconstructed events is about 2\,600. The $\Delta m_B$ value
retrieved in the cross-check procedure explained above is in good agreement with the world average, making the $w$ value
trustable.
Thanks to this high number of reconstructed events, the effect of the incertainty on the measured $w$ value on the \sb\ measurement
is kept smaller than the statistical error on \sb .

\subsection{Dilepton mixing}

This analysis is quite similar to the previous one except that only leptons are considered to infer the \B\ flavor.
It measures the asymmetry between ``same sign'' and ``opposite sign'' di-lepton events:

\[ A(\Delta t) = \frac{N_B(l^\pm l^\mp) - N_B(l^\pm l^\pm)}{N_B(l^\pm l^\mp) + N_B(l^\pm l^\pm)} \]

\begin{figure}[!htb]
\begin{center}
\includegraphics[height=5.0cm]{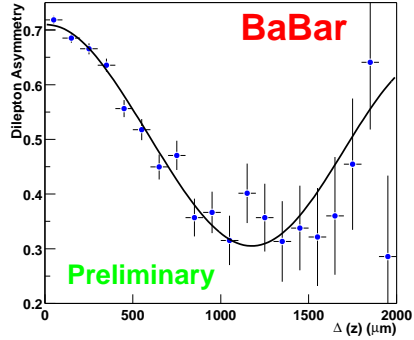}
\caption{Dilepton $l^\pm l^\mp$ - $l^\pm l^\pm$ asymmetry.}
\label{fig:dilasym}
\end{center}
\end{figure}

The today observed asymmetry is shown on figure ~\ref{fig:dilasym}.
This analysis should provide a competitive measurement on $\Delta m_B$.

\section{Conclusion}
\babar\ and \pep2\ have had a terrific start. 11 \invfb\ have been recorded since the end of May 1999
and the \babar\ detector performances are close to the design values.

The ingredients necessary to the \sb\ measurement are under control: the ``golden plated'' channel \bpsiks\ is reconstructed with low
background and understood yield. The $\Delta z$ measurement and the tagging algorithm are controled on the data themselves.

The \babar/\pep2\ short and medium term plans are to run up to the end of October, with an expected collected data
sample larger than
20 \invfb\ and then to run at the designed luminosity $3\times 10^{33}\cms$, which should allow \babar\ 
to collect 30 \invfb\ in one year.\\


\section{Acknowledgments}
\label{sec:Acknowledgments}

We are grateful for the contributions of our \pep2\ colleagues in
achieving the excellent luminosity and machine conditions
that have made this work possible.
We acknowledge support from the
Natural Sciences and Engineering Research Council (Canada),
Institute of High Energy Physics (China),
Commissariat \`a l'Energie Atomique and
Institut National de Physique Nucl\'eaire et de Physique des Particules
(France),
Bundesministerium f\"ur Bildung und Forschung
(Germany),
Istituto Nazionale di Fisica Nucleare (Italy),
The Research Council of Norway,
Ministry of Science and Technology of the Russian Federation,
Particle Physics and Astronomy Research Council (United Kingdom), the
Department of Energy (US),
and the National Science Foundation (US). In addition, individual support 
has been received from the Swiss 
National Foundation, the A. P. Sloan Foundation, the Research Corporation,
and the Alexander von Humboldt Foundation.
The visiting groups wish to thank 
SLAC for the support and kind hospitality
extended to them.

\end{document}